\documentclass[fleqn,usenatbib]{mnras}

\usepackage[T1]{fontenc}
\usepackage{ae,aecompl}

\usepackage{graphicx}	% Including figure files
\usepackage{amsmath}	% Advanced maths commands
\usepackage{amssymb}	% Extra maths symbols

\title[The M-$\sigma$ relation from the disruption of binaries]{The M-$\sigma$ relation from the disruption of binaries from the galactic bulge}

\author[E. Michaely and D. Hamilton]{
Erez Michaely\thanks{E-mail: erezmichaely@gmail.com} and
Douglas Hamilton
\\
Astronomy Department, University of Maryland, College Park, MD 20742}

\date{Accepted XXX. Received YYY; in original form ZZZ}

\pubyear{2020}

\begin{document}
\label{firstpage}
\pagerange{\pageref{firstpage}--\pageref{lastpage}}

\maketitle

\begin{abstract}
We present a novel explanation of the well known $M_{\bullet}-\sigma$
relation. In a triaxial potential binaries with chaotic orbits within
a sphere that encompass $\sim100$ times the mass of the super-massive
black-hole (SMBH) have a finite probability to be tidally disrupted
by the SMBH. As a result one component loses energy and might itself
break apart tidally and accreted onto the SMBH. More significantly,
the other component, which gains energy, returns to the bulge and
equilibrates its excess energy with the environment thereby changing
the kinetic temperature, hence the velocity dispersion. We develop
a mathematical model and find that its results are in agreement with
the observed relation.
\end{abstract}

\section{Introduction}
\label{sec:Introduction}
The mass of a super massive black-hole (SMBH), $M_{\bullet}$, is
correlated with several properties of its host galaxy. $M_{\bullet}$
has correlations with the stellar luminosity \citep{Gueltekin2009},
$M_{\bullet}-L_{*}$ relation; with stellar mass, specifically the
mass of the bulge (for disk galaxies) or the mass of the galaxy itself
(for elliptical galaxies) \citep{McConnell2013}. Surprisingly, the
tightest relation is with the stellar velocity dispersion of the spheroid
surrounding the SMBH, the well known $M_{\bullet}-\sigma$ relation
\citep{Ferrarese2000,Zubovas2019,Gueltekin2009}. These relations
provide important evidence for the co-evolution of both the SMBH and
the host galaxy which have interesting ramifications in many fields
of astrophysics. In this work we focus on the $M_{\bullet}-\sigma$
relation, namely, $\log M_{\bullet}=\gamma+\beta\log\sigma$.

The first $M_{\bullet}-\sigma$ relation observational papers were
published almost 20 years ago \citep{Ferrarese2000,Gebhardt2000}.
Both teams reported their results with almost no scatter, but with
different slopes. \citet{Ferrarese2000} reported $\beta=4.80\pm0.50$
while \citet{Gebhardt2000} found $\beta=3.75\pm0.30$. In course
of time the $M_{\bullet}-\sigma$ relation was measured for more galaxies.
\citet{vandenBosch2016} calculated $\beta=5.35\pm0.23$ and $\gamma=-4.0\pm0.5$;
while \citet{McConnell2013} got $\beta=5.64\pm0.32$ and $\gamma=8.21\pm0.05$.
Regardless of the precise value, the $M_{\bullet}-\sigma$ relation
is both remarkable and surprising. The SMBH dominates gravitationally
the immediate vicinity of its location, the sphere of influence, which
has typically radius of a few parsecs. However, the radius of the
spheroids surrounding the SMBH have typical radii of a kilo parsec
(or more), so the SMBH cannot govern the dynamics of it. Yet the $M_{\bullet}-\sigma$
relation is observed in the local Universe, at redshifts $z\lesssim0.1$,
corresponding to time later than $t\sim12{\rm Gyr}$ after the Big
Bang.

\citet{Shen2015} used data from the Sloan Digital Sky Survey in order
to search for the $M_{\bullet}-\sigma$ relation as a function of
redshift. They report that no evidence of such evolution and the $M_{\bullet}-\sigma$
relation holds up to $z\approx1$, where the age of the Universe at
$z\approx1$ is $\sim6{\rm Gyr}.$ This suggests that the process
that leads to the $M_{\bullet}-\sigma$ relation should saturate by
$t\sim6{\rm Gyr}$.

Generally speaking the proposed explanations for the $M_{\bullet}-\sigma$
relation could be grouped into three categories \citep{Zubovas2019}.
First, ``central limit-like theorem'', \citep[e.g.][]{Peng2007,Jahnke2011}.
In these sets of explanations, the underlining assumption is that
the $M_{\bullet}-\sigma$ relation emerges not due to co-evolution
of the central BH and the host galaxy, but due to hierarchical assembly
of BH and stellar mass through galaxy mergers. The mergers from an
initially uncorrelated distribution of BH and stellar masses in the
early universe produce the observed correlations. Second, ``Gas feed
rate'' \citep[e.g.][]{Haan2009,Angles-Alcazar2013,Angles-Alcazar2015}.
This theory proposes that the SMBH mass growth is due to the feeding
of gas which in turn is a function of the host galaxy characteristics,
specifically galaxy-scale torques on the gas that govern the inflow
of gas to the SMBH and hence govern the mass of the SMBH. Third, and
arguably the most accepted explanation is the ``Feedback mechanism''.
This process relies on the energy released from the accretion of mass
on to the SMBH. The energy released may affect the entire galaxy which
can regulate, in turn, the mass infall to the SMBH. Feedback can come
in several forms, changing the star formation rate or regulating the
infall mass rate itself onto the SMBH.

Almost 20 years ago Merrit and Poon published a series of four papers
on triaxial nuclear bulges containing an SMBH \citep[hereafter PM1;PM2;PM3;MP4]{Poon2001,Poon2002,Poon2004,Merritt2004}.
In PM1 they investigated the orbital motion of test particles in a
triaxial nucleus hosting an SMBH. The stellar density profile they
consider follows a power law $\rho_{*}\propto r^{-\gamma}$ with $\gamma=\left\{ 1,2\right\} $.
For triaxial potentials with a central point mass the phase space
is naturally divided into three regions defined by the distance (energy)
from the center. The innermost region, within the sphere of influence
of the SMBH, with radius $r_{h}\approx GM_{\bullet}/\sigma^{2}$,
hosts low energy orbits, e.g. tubes, pyramids and bananas, and the
trajectories are mainly regular and avoid close passages with the
center of the potential. However, at higher energies the pyramid orbits
become increasingly chaotic. The transition to the chaotic regime
occurs rapidly, i.e. sharply in space. Beyond the sphere of influence
is the second region with intermediate radii, the scattering zone
region. In this region, the SMBH acts as a scattering center for almost
all the center-filling trajectories. The second region is located
from the edge of the sphere of influence outward until the radius
that encompass a total mass of $\sim50-100M_{\bullet}$. In this region
the orbits are a mix of ``regular'' orbits which avoid the center
of potential and ``chaotic'' orbits which pass near the center of
potential one per crossing time. The fraction of chaotic orbits is
$f_{c}\approx0.5$ (MP4). The third region, the outermost region,
hosts the highest orbital energies, and the rest of the mass of the
spheroid. The phase space is a complex mixture of chaotic and regular
trajectories. This region has a mixture of chaotic and regular orbits.

In PM2 they showed that the triaxial potential is retained in time.
Hence one cannot overlook the importance of stellar dynamics in the
environments of triaxial galactic potentials. PM3 investigated the
fraction of chaotic orbits for 3 triaxial shapes: almost prolate,
almost oblate and maximally triaxial. They found that $\sim50\%$
of the mass is assigned to chaotic orbits. The last paper of the series,
MP4, present a mathematical model of the galactic center and calculates
the rate of single-star disruption from chaotic orbits in order to
explain the $M_{\bullet}-\sigma$ relation.

In this paper we build on the work of MP4 and expand their modeling
to binaries that are tidally disrupted by the SMBH. In what follows,
we describe the co-evolution of the SMBH mass growth together with
the change in the kinetic temperature of the spheroid due to disruption
of binaries from the bulge. As a result of the binary disruption,
a fraction of single stars will experience a stellar tidal disruption
event (TDE), while the surviving star re-equilibrates its excess with
the bulge altering its kinetic temperature and hence the velocity
dispersion.

In section \ref{sec:The-Model} we describe the model both qualitatively
and quantitatively. In section \ref{sec:Results} we present the results
of the numerical simulation while in section \ref{sec:Discussion-and-Summary}
we discuss implications and caveats and summarize the manuscript.

\section{The Model}
\label{sec:The-Model}
\subsection{Qualitative description}

In this subsection we briefly describe the dynamical model and assumptions
in a qualitative manner. We assume triaxial potentials for all bulges
with isotropic mass distributions. The number of systems (either binaries
or single stars) is $N=N_{b}+N_{s}$ where $N_{b\left(s\right)}$
is the number of binaries (singles). Furthermore, we assume that the
initial binary fractions equal to $f_{{\rm binary}}$, i.e. the mass
in binaries is $f_{{\rm binary}}\times M_{{\rm bulge}}$. For simplicity
we set all binaries components to have the same mass $m_{1}=m_{2}=1M_{\odot}$
in circular orbits. The semi-major axis (sma) is distributed from
some distribution function $f_{a}$. PM1-3 and MP4 showed that the
centrophilic orbits are about half of the stellar mass of the second
spatial region, hereafter bulge mass, i.e. half of the binaries. In
our model we calculate the rate that binaries enter the binary tidal
disruption radius, $r_{{\rm bu}}$. As a result, from this binary
disruption one component returns to the bulge with typically more
specific energy, hence the energy budget of the bulge changes and
so the velocity dispersion evolves. The other component, which is
typically captured/disrupted by the SMBH, may change the mass of the
SMBH. Additionally, we account for the binary ionization process in
the bulge, due to random interaction with passing stars.

We model the binary tidal disruption with the impulse approximation.
The impulse approximation holds when the binary may be considered
satationary while interacting with the SMBH. The two relevant timescales
are the binary orbital period, $P$ and the interaction timescale,
$t_{{\rm int}}\equiv q/v_{q}$ where $q$ is the closest approach
of the binary center of mass to the SMBH and $v_{q}$ is the center
of mass velocity at $q$ \citep{Agnor2006}.

In order to verify the validity of the impulse approximation for a
binary interaction with the SMBH, we perform a set of $1000$ numerical
simulations. Using an N-body integrator \citep{Hut1981} we simulate
a circular binary with two component masses of $m_{1}=m_{2}=1M_{\odot}$,
with a center of mass on a hyperbolic trajectory around an SMBH with
mass of $M_{\bullet}=4\times10^{6}M_{\odot}.$ We initiate all binary
center of mass velocities to be equal to the bulge velocity dispersion,
namely $\sigma=200{\rm kms^{-1}}$. Next we set the binary semi-major
axis (sma), $a$, the pericenter distance, to the SMBH, $q$, and
the binary mean anomaly, ${\scriptscriptstyle M}.$ We sample 10 equally
spaced sma values in log space between $10^{-2}{\rm AU}$ and $10^{2}{\rm AU}$.
Additionally, we set $10$ equally spaced pericenter values in log
space between $r_{{\rm *}}$ and $r_{{\rm bu}}$, where $r_{*}$ is
the tidal disruption radius of a single star, given by (\ref{eq:star_tidal})
and $r_{{\rm bu}}$ is the binary tidal disruption radius, given by
(\ref{eq:binary_tidal}); where $m_{*}$ and $R_{*}$ is the mass
and radius of a star and $m_{b}=m_{1}+m_{2}$ is the total mass of
the binary system.
\begin{equation}
r_{*}=\left(\frac{3M_{\bullet}}{m_{*}}\right)^{1/3}R_{*}\label{eq:star_tidal}
\end{equation}

\begin{equation}
r_{{\rm bu}}=\left(\frac{3M_{\bullet}}{m_{b}}\right)^{1/3}a\label{eq:binary_tidal}
\end{equation}
Furthermore, we set $10$ equally spaced mean anomaly values, ${\scriptscriptstyle M}$
between $0$ and $\pi.$ We align the binary angular momentum vector
with the angular momentum of the hyperbolic trajectory, i.e. setting
the inclination to zero, for simplicity.

For each simulation that underwent binary disruption, we record the
outcomes of the components, namely one component is always ejected
and the other is either ejected or captured/disrupted. We emphasize
that the initial trajectory is hyperbolic hence both components may
escape after the fly-by. Next we focus on the component with the higher
kinetic energy, we calculate its velocity at the edge of the Hill
sphere, $r_{h}$. Using conservation of energy where $v_{{\rm bu}}$
is the circular velocity around the binary center of mass: 
\begin{equation}
v_{b}=\left(2GM_{\bullet}\left(\frac{1}{r_{{\rm h}}}-\frac{1}{r_{{\rm bu}}}\right)+v_{{\rm bu}}^{2}\right)^{1/2}.\label{eq:v_at_the_bulge}
\end{equation}

\begin{figure}
\includegraphics[width=0.9\columnwidth]{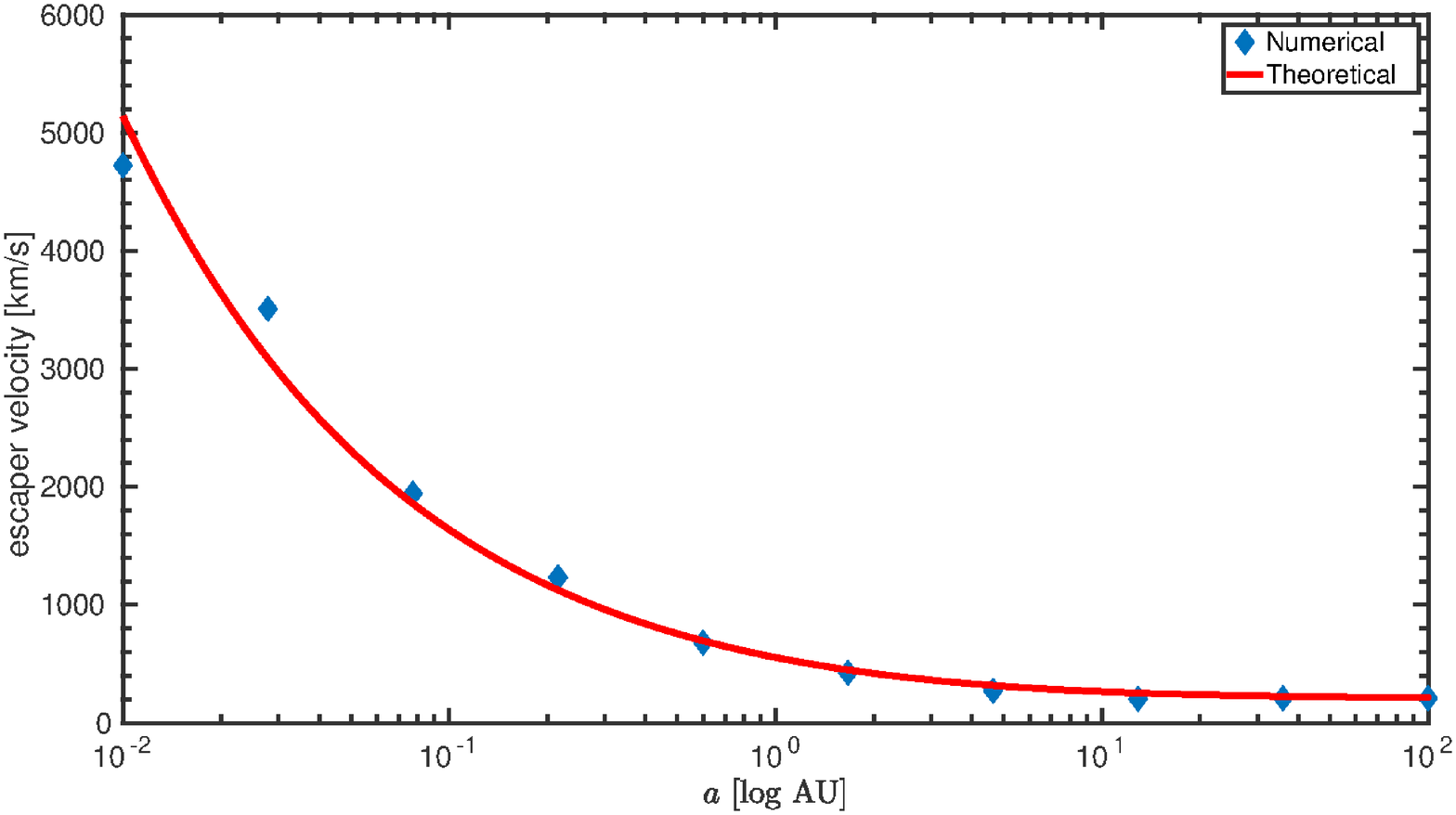}\caption{\label{fig:impulse_approx}Verification of the impulse approximation
treatment. The value of the escaper's velocity at the edge of the
radius of influence as a function of binary sma. The blue dimonds
are the calculated velocity of the escaper at the edge of the sphere
of inluence, from the N-body simulation. The red solid line is the
predicted values of the velocity from the theoretical treatment of
the impulse approximation. The agreement is good.}
\end{figure}

Figure \ref{fig:impulse_approx} presents the velocity of the escaper
at the edge of the Hill sphere as a function of initial binary sma.

Next we focus on the component with the lower kinetic energy.

\subsection*{The less energetic component}

\begin{figure}
\includegraphics[width=0.9\columnwidth]{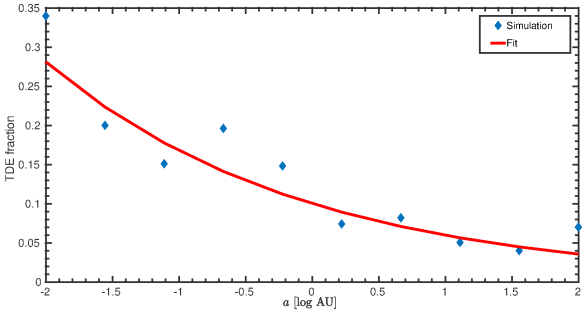}\caption{\label{fig:TDE_fraction}Blue dimonds are the fraction of TDE as a
function of binary sma from the numerical simulation. Red solid line
is the best fit $f_{{\rm TDE}}=0.1\times\left(\frac{a}{{\rm AU}}\right)^{-0.2244}$
for $M_{\bullet}=1\times10^{6}M_{\odot}$ and $\sigma=200{\rm kms^{-1}}.$}
\end{figure}

In the previous subsection we focused on the more energetic component,
and approximated its velocity when reaching the bulge boundary, i.e.
the edge of the sphere of influence. In this subsection we focus on
the less energetic component. This component acquires a new Keplerian
trajectory upon binary disruption. We record its closest approach
to the SMBH and compare it to $r_{*}$, the tidal disruption radius
of a single star. If the closest approach is smaller than $r_{*}$
than we flag it as a tidal disruption event (TDE). Figure \ref{fig:TDE_fraction}
shows the fraction of single-star TDE out of the disrupted binary
sample as a function of binary sma from the simulation. We found the
best fit for the TDE fraction as a function of binary sma, $a$, to
be the following:
\begin{equation}
f_{{\rm TDE}}\left(a\right)=0.1\times\left(\frac{a}{{\rm AU}}\right)^{-0.2244}.\label{eq:f_TDE}
\end{equation}

\subsection{Quantitative description}

\subsubsection*{Mathematical framework}

In this subsection we describe in detail the mathematical model accounting
for the physical processes.

There are three relevant length scales in the problem regarding the
binary break-up and the tidal distruption of one of the stars from
the binary. First, $r_{{\rm bu}}$, is the binary tidal disruption
radius from the SMBH for a binary with sma $a$ and total binary mass
$m_{b}$, as defined in equation (\ref{eq:binary_tidal}). For an
SMBH mass of $M_{\bullet}=10^{6}M_{\odot}$ and a binary with sma
of $a=1{\rm AU}$ and total mass of $m_{b}=2M_{\odot}$, the value
of binary tidal disruption radius is of the order of $r_{{\rm bu}}\approx100{\rm AU}.$
Second, $q$, is the pericenter distance of the binary trajectory
from the SMBH. Third, $r_{*}$, is the single star tidal disruption
radius for a star with radius $R_{*}$ and mass $m_{*}$ defined in
equation (\ref{eq:star_tidal}). For a sun like star the value of
the tidal disruption radius is approximatly $r_{*}\approx1{\rm AU}.$

The bulge is naturally divided into three spatial regions for a triaxial
potential with a central point mass (the SMBH) as mentioned in section
\ref{sec:Introduction}. In this work we focus on the second region,
the scattering zone, where roughly half of the mass are in chaotic
trajectories. The total stellar mass in it is $\sim100M_{\bullet}$.

We use the model and notations for the galactic nucleus described
in MP4. The stellar density is given by
\begin{equation}
\rho_{*}=\rho_{0}m^{-\gamma}\label{eq:density_r}
\end{equation}
we get set $\rho_{0}=1$ because of the scale free nature of the density
profile with no loss of generality, $m$ is defined by the following
equation of an ellipsoid:

\begin{equation}
m^{2}=\frac{x^{2}}{a^{2}}+\frac{y^{2}}{b^{2}}+\frac{z^{2}}{c^{2}}.
\end{equation}
 The triaxial parameter $T$ is given by

\begin{equation}
T\equiv\frac{a^{2}-b^{2}}{a^{2}-c^{2}}
\end{equation}
 where $T=0.5$ is maximally triaxial.

We focus on the case where $\gamma=2$, the isothermal sphere with
particle mass of $m_{b}$. In this steep cusp profile the potential
is given by equation 7 in PM1. It is convenient to use the corresponding
circular orbit energy in the analogous spherical model is 
\begin{equation}
E_{c}\left(r\right)=4\pi\delta^{2}\left[\ln\left(\frac{r}{\delta}\right)-1\right]\label{eq:energy_circular}
\end{equation}
where $\delta=\left(abc\right)^{1/3}=0.734$ for $T=0.5$. This function
is given in model units specified in MP4. In this system the units
of mass, length and time is given by the following: 
\begin{equation}
\left[M\right]=M_{\bullet}\qquad\left[L\right]=\left(2\pi\delta^{2}\right)r_{h}\qquad\left[T\right]=\left(2\pi\delta^{2}\right)^{3/2}\sqrt{\frac{r_{h}^{3}}{GM_{\bullet}}}.
\end{equation}
Here we follow MP4 to define $r_{h}=GM_{\bullet}/\sigma^{2}$ as the
radius in the spherical model containing a stellar mass of $2M_{\bullet}$.

Given equations (\ref{eq:density_r}) and (\ref{eq:energy_circular})
we write the total mass of stars per specific energy in real units:
\begin{equation}
\mathcal{M}\left(E\right)=\frac{2\sqrt{6}}{9}\frac{M_{\bullet}}{\sigma^{2}}\exp\left(\frac{E-E_{h}}{2\sigma^{2}}\right)\label{eq:Mass_with_E}
\end{equation}
where $E_{h}\equiv E\left(r_{h}\right)$. We define $\mathcal{M}_{b}\left(E\right)\equiv\left(N_{b}/N\right)\mathcal{M}\left(E\right)$
to be the total mass in binaries with specific energy shell $E.$
The corresponding density as a function of distance in model units
and real units is:
\begin{equation}
\rho_{*}=\rho_{h}\left(\frac{r}{r_{h}}\right)^{-2}=\frac{M_{\bullet}}{2\pi r_{h}}r^{-2}.
\end{equation}
where we have chosen $\gamma=2$ in equation \ref{eq:density_r} with
the normalization $\rho_{h}=M_{\bullet}/\left(2\pi r_{h}^{3}\right)$
which is the stellar density at the edge of the sphere of influence.

MP4 present their results on the number of encounters per unit time
for a chaotic orbit, with some energy $E$, within a distance $d$
from the center of the potential, i.e. the SMBH. They found a linear
scaling with $d$ combined with the gravitational focusing from the
SMBH, where the cross section scales linearly with $d$ as well, the
number of encounters within a distance $d$ per unit time scales like
$N_{<d}\propto d^{2}$ for a given energy shell. Generally they found
the rate per unit time per unit distance is 
\begin{equation}
A\left(E\right)\approx1.2\left(\frac{\sigma^{5}}{\left(GM_{\bullet}\right)^{2}}\right)\exp\left(\frac{-\left(E-E_{h}\right)}{\sigma^{2}}\right).\label{eq:rate_per_unit_length}
\end{equation}

\subsubsection*{Binary disruption calculation}

The rate at which a binary on a chaotic orbit of specific energy $E$
experiences closest approach to the SMBH within a distance $d$ is
given by $A\left(E\right)\times d$. Together with equation (\ref{eq:Mass_with_E})
we can write the rate of binary disruption as a function of time,
$t,$sma, $a$ and energy, $E$, by setting $d=r_{{\rm bu}}$ 
\begin{equation}
\Gamma_{1}\left(a,E,t\right)=f_{a}\left(a\right)A\left(E\right)r_{{\rm bu}}\left(a\right)\frac{f_{c}M\left(E\right)}{m_{b}}\exp\left(-A\left(E\right)r_{{\rm bu}}\left(a\right)t\right).\label{eq:gamma_no_ion}
\end{equation}
We remind that $f_{c}M\left(E\right)/m_{b}$ is the number of binaries
in chaotic orbits and $f_{a}\left(a\right)$ is the sma distribution.
This equation does not account for binary ionization in the bulge
due to random interaction with passing systems. These interactions
may disrupt binaries resulting in a reduction of the number of available
binaries to be disrupted by the SMBH. In order to account for this
we calculate the half-life time of a binary with sma $a$ and total
mass $m_{b}$ in an environment with stellar density $\rho_{*}$ and
velocity dispersion $\sigma$ by \citep{Bahcall1985}
\begin{equation}
t_{1/2}\left(a,E\right)=0.00233\frac{\sigma}{G\rho_{*}\left(E\right)a}.
\end{equation}
As a result the corrected binary disruption rate is
\begin{equation}
\Gamma\left(a,E,t\right)=\Gamma_{1}\left(a,E,t\right)\exp\left(-t\ln2/t_{1/2}\left(a,E\right)\right).\label{eq:ionization_rate}
\end{equation}

\subsubsection*{Implications from binary disruption}

Once a binary, in a hyperbolic trajectory, enters $r_{{\rm bu}}$
it is disrupted into its two components stars. One star receives energy
and return to the bulge, $m_{1}$, while the other star with mass
$m_{2}$ may be capture or disrupted by the SMBH. As a result the
mass of the bulge is reduced by $m_{2}$, while the mass of the SMBH
increases by the amount of the accreted mass from a possible single
star TDE, $\Delta m_{{\rm acc}}$. We define $\Delta m_{{\rm acc}}\equiv m_{2}\times f_{{\rm TDE}}\left(a\right)\times f_{{\rm acc}}$
where $f_{{\rm TDE}}$ is the fraction of TDE from the set of binary
tidal disruptions, and $f_{{\rm acc}}=1/2$ (PM3) for $M_{\bullet}<10^{8}M_{\odot}$
and $f_{{\rm acc}}=1$ for $M_{\bullet}>10^{8}M_{\odot}.$

Additionally, once $m_{1}$ returns back to the bulge it arrives with
speed $v_{1}\left(a\right)$. The bulge is modeled as an isothermal
sphere with Maxwellian distribution of velocities. In this case a
kinetic temperature of the bulge can be defined as a function of the
velocity dispersion:

\begin{equation}
k_{B}T_{{\rm bulge}}=\bar{m}\sigma^{2}=\frac{\bar{m}\bar{v^{2}}}{3}
\end{equation}
where $\bar{m}$ is the average mass of the components in the bulge,
$\bar{v^{2}}$ is the mean square speed of the components and $T_{{\rm bulge}}$
is the kinetic temperature of the bulge. Therefore, one can determine
the equivalent kinetic temperature of a single star to be 
\begin{equation}
T_{2}=\frac{m_{1}v_{1}^{2}\left(a\right)}{k_{B}}.
\end{equation}

In this work we assume $m_{1}$ equilibrates its energy with the environment
of the bulge, we address this assumption in section \ref{sec:Discussion-and-Summary}.
The change in bulge temperature per unit time, due to this process
is 
\begin{equation}
\frac{dT_{{\rm bu}}}{dt}=\left(\frac{1}{N}\right)\int_{a_{{\rm min}}}^{a_{{\rm max}}}da\int_{{\rm E_{h}}}^{E_{{\rm edge}}}dE\times\Gamma\left(a,E\right)\times\left(T_{2}\left(a\right)-T_{{\rm bulge}}\right).
\end{equation}

Where $E_{{\rm edge}}$ is the energy at the end of the second spatial
region and $a_{{\rm min}},$ $a_{{\rm max}}$ are the boundaries of
the sma. Moreover, binaries in the bulge get disrupted continuously
due to random interaction with passing stars. These process changes
both the number of components (from a binary to two single stars)
and the specific energy of the bulge. The two stars come with kinetic
temperature of 
\begin{equation}
T_{3}\left(a\right)=\frac{2m_{*}}{k_{B}}\left(\frac{1}{2}\sqrt{\frac{Gm_{b}}{a}}\right)^{2}.
\end{equation}
The rate where $N_{b}\left(t\right)$ decreases both due to the ionization
process and the binary disruption from the SMBH is given by the following:

\[
\frac{dN_{b}\left(t\right)}{dt}=-\int_{a_{{\rm min}}}^{a_{{\rm max}}}da\int_{{\rm E_{h}}}^{E_{{\rm edge}}}dE\times
\]
\begin{equation}
\left(\frac{\mathcal{M}_{b}\left(E\right)}{m_{b}}f_{a}\left(a\right)\frac{\ln2}{t_{1/2}\left(a,E\right)}+\Gamma\left(a,E,t\right)\right).\label{eq:N_binaries_rate}
\end{equation}

We calculate the change in temperature due to this process by
\begin{equation}
\frac{dT_{{\rm ion}}}{dt}=\int_{a_{{\rm min}}}^{a_{{\rm max}}}da\int_{{\rm E_{h}}}^{E_{{\rm edge}}}dE\frac{1}{N}\frac{dN_{b}\left(t\right)}{dt}\left(T_{3}\left(a\right)-T_{{\rm bulge}}\right),
\end{equation}
which together with 
\begin{equation}
\sigma\left(t\right)=\text{\ensuremath{\left(\frac{k_{B}T\left(t\right)}{\frac{1}{N}\left(N_{{\rm b}}\left(t\right)\times m_{b}+N_{{\rm s}}\left(t\right)\times m_{*}\right)}\right)}}^{1/2}
\end{equation}
allows us to calculate the evolution of the velocity dispersion.

\subsubsection*{Mass accretion to the SMBH}

The mass accretion to the SMBH originates from TDEs after binaries
are disrupted whence their pericenter distances are closer than $r_{{\rm bu}}.$
Some fraction from all disrupted binaries ends up with a single star
TDE, $f_{{\rm TDE}}\left(a\right)$. In this work we assume that half
of the mass of the disrupted star is accreted \citep{Stone2019a}
for $M_{\bullet}<10^{8}M_{\odot}$ and all of the mass of the disrupted
star is accreted for $M_{\bullet}>10^{8}M_{\odot}$. 
\begin{equation}
\frac{dM_{\bullet}\left(a,E\right)}{dt}=f_{{\rm TDE}}\left(a\right)\times\Gamma\left(a,E\right)\times m_{*}.
\end{equation}

\section{Results}
\label{sec:Results}
In this section we present results for a wide range of plausible initial
conditions and binaries properties in order to show the robustness
of the proposed process. In subsection \ref{subsec:Time-evolution:-Representative}
we show a representative example of the time evolution of some model
parameters. In subsection \ref{subsec:Main-results} we describe the
initial conditions used and present the evolution of the $M_{\bullet}-\sigma$
relation for the considered calculation.

\subsection{Time evolution: Representative example}
\label{subsec:Time-evolution:-Representative}
\begin{figure}
\includegraphics[width=0.90\columnwidth]{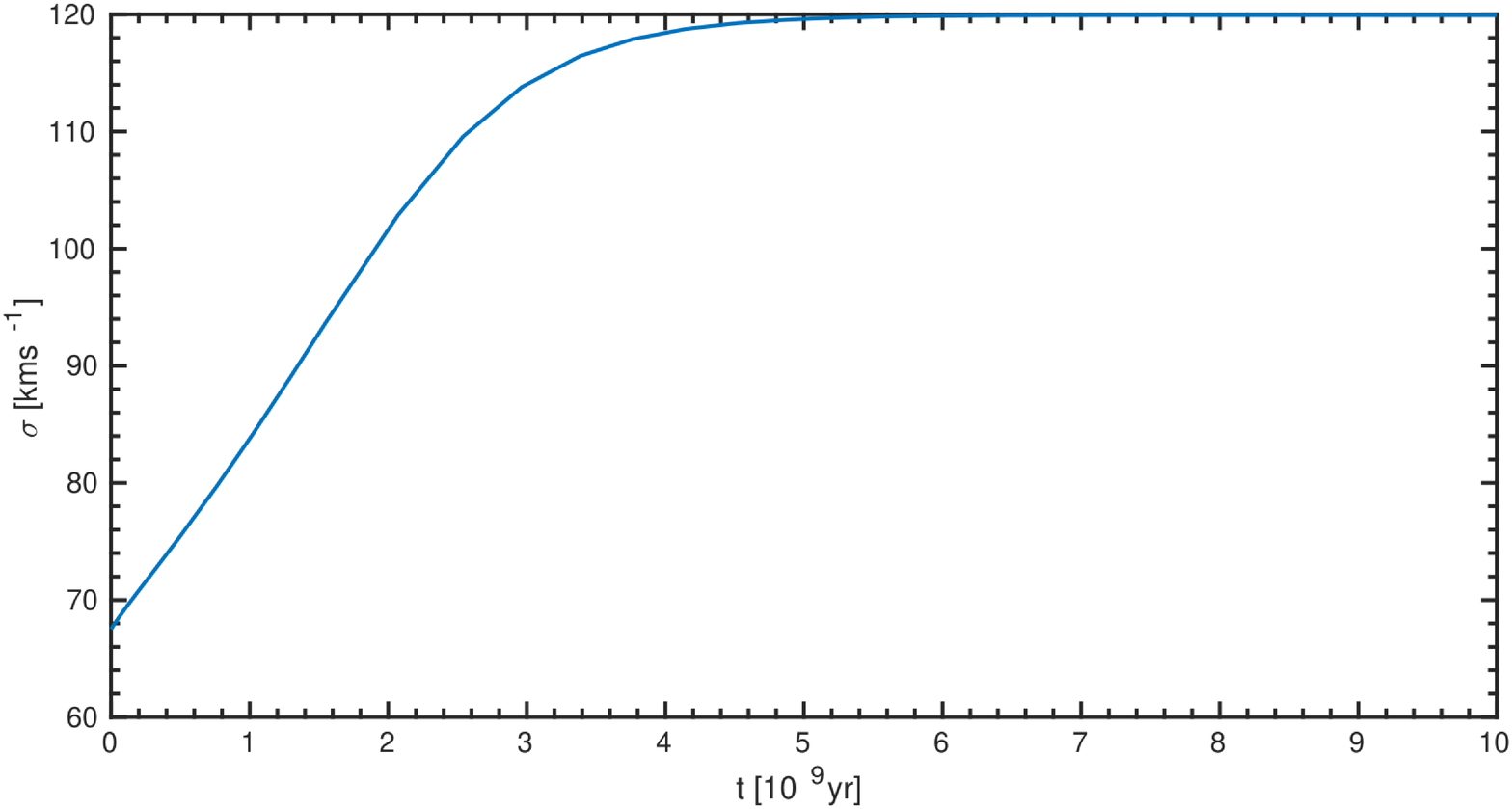}

\includegraphics[width=0.90\columnwidth]{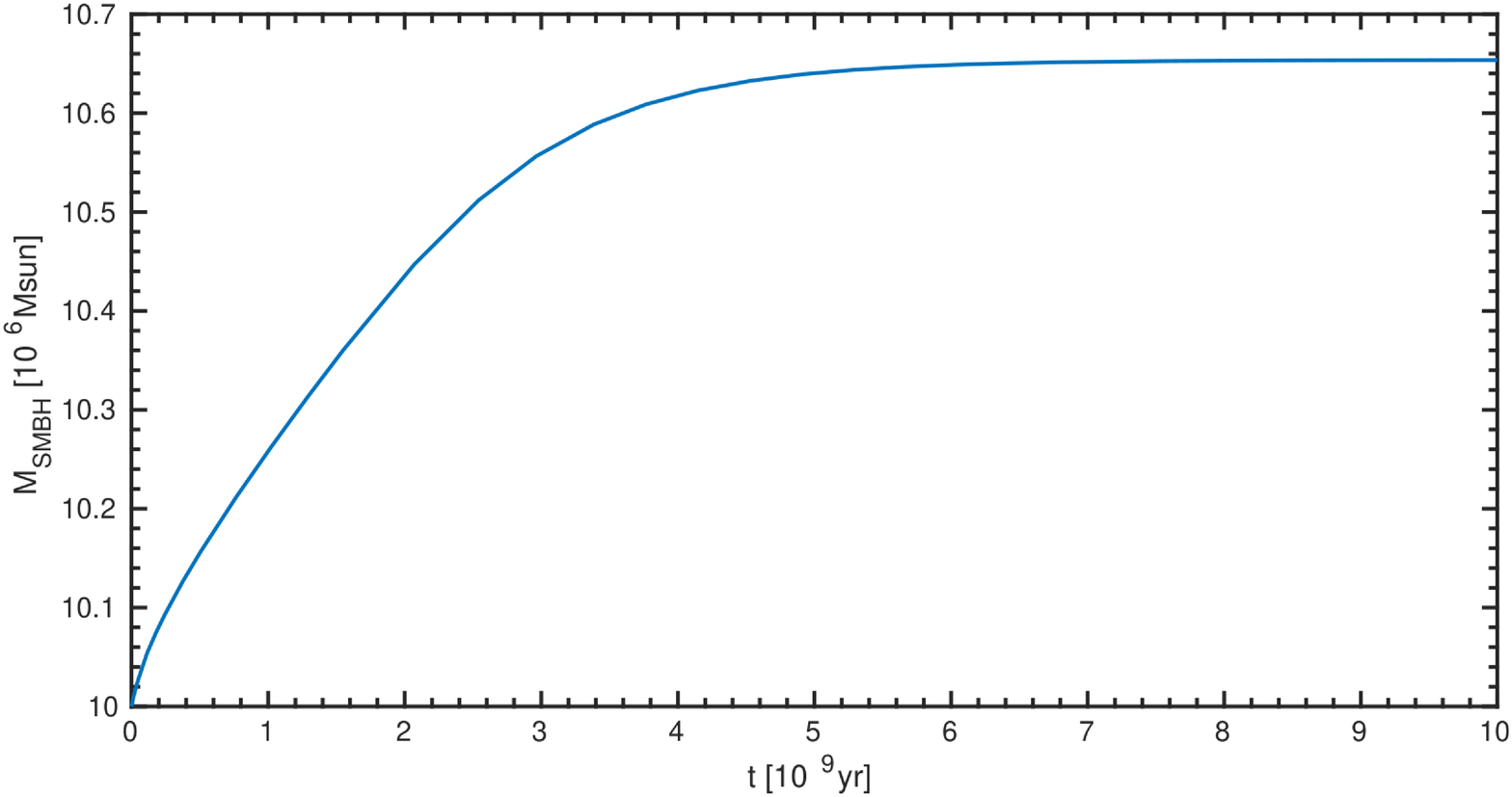}

\includegraphics[width=0.90\columnwidth]{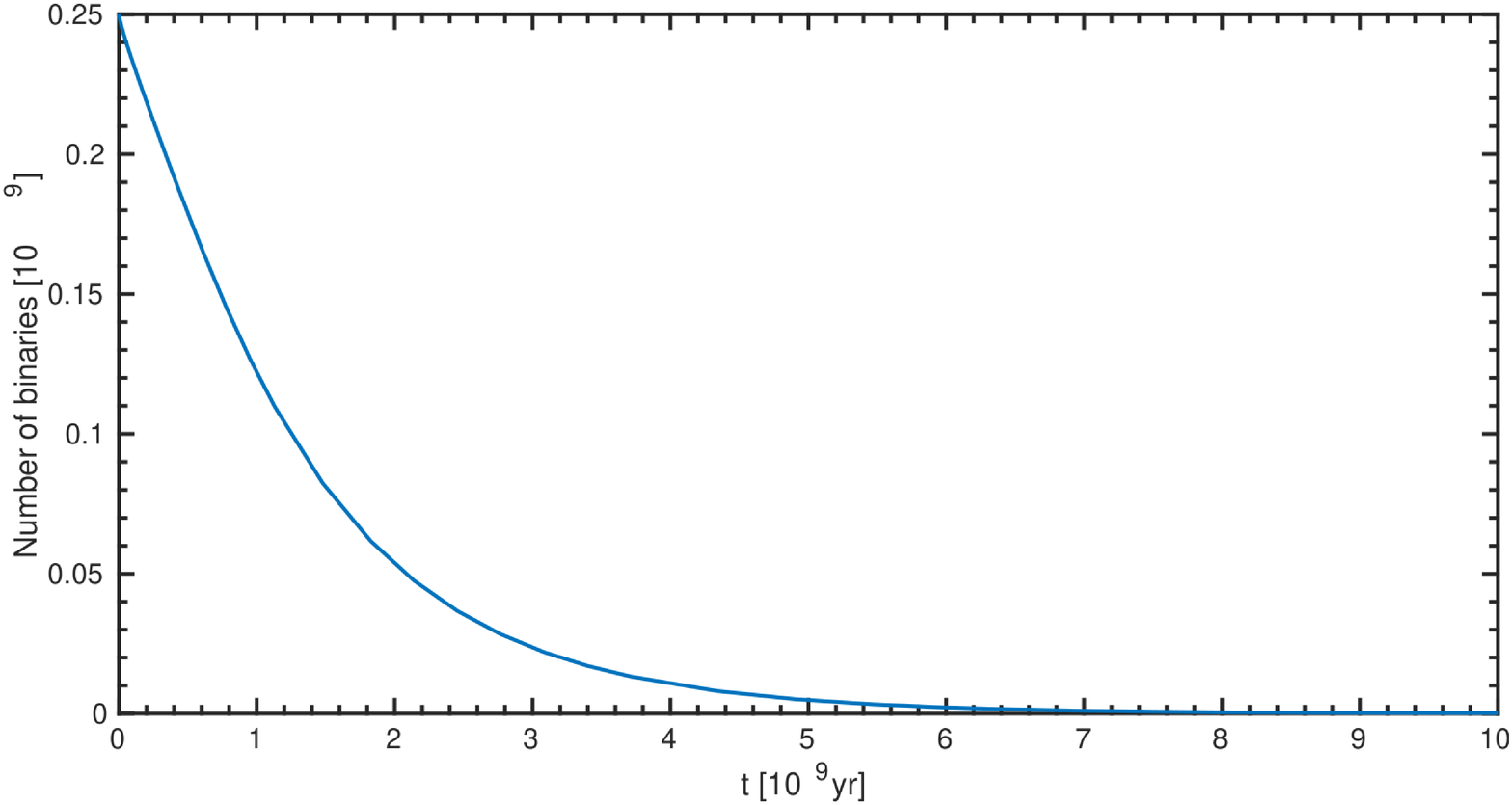}

\includegraphics[width=0.90\columnwidth]{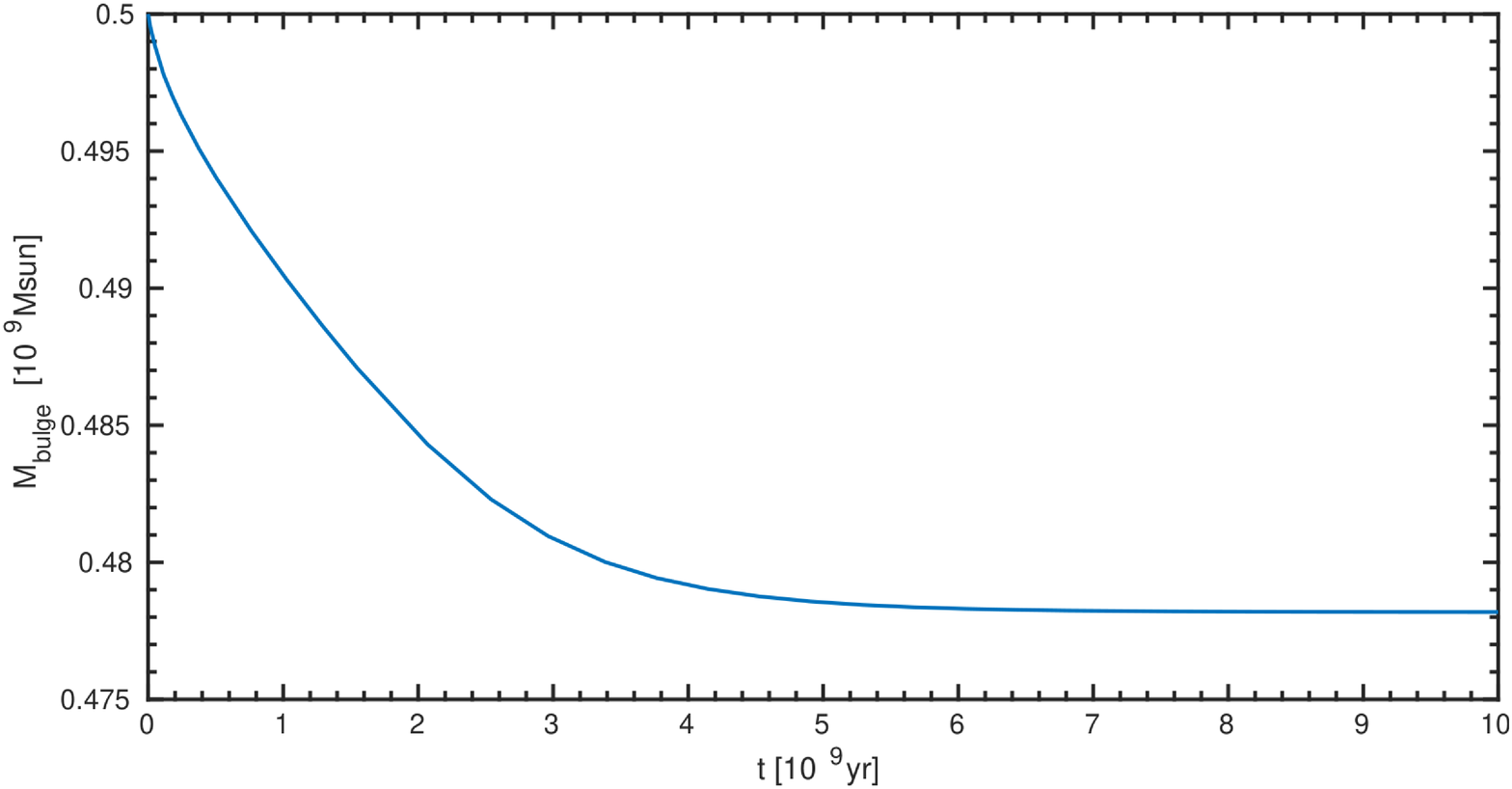}\caption{\label{fig:Example_time_evolution}The time evolution the velocity
dispersion and the SMBH mass. For $\alpha\equiv M_{{\rm bulge}}/M_{\bullet}=50$,
$M_{\bullet}=1\times10^{7}{\rm M_{\odot}}$, $\sigma_{0}=67.5{\rm kms^{-1}}$,
$a_{{\rm min}}=0.01{\rm AU},$ $a_{{\rm max}}=100{\rm AU},$ log uniform,
$f_{{\rm TDE}}=0.1\times\left(\frac{a}{{\rm AU}}\right)^{-0.2244}.$
The binary fraction used here was $f_{{\rm binary}}=1.$ The upper
plot presents the evolution of the velocity dispersion $\sigma$
in time. The second from the top presents the change in the SMBH mass
as a function of time. The third from the top shows the decrease in
the numbers of binaries in the bulge due to both ionization and binary
disruption by the SMBH (\ref{eq:N_binaries_rate}). The bottom
panel presents the change in the mass of the bulge.}
\end{figure}

For the representative example we consider an SMBH with initial mass
of $M_{\bullet}=10^{7}{\rm M_{\odot}}$ embedded in a bulge with $M_{{\rm bulge}}=5\times10^{8}{\rm M_{\odot}}$.
We emphasize that the definition of $M_{{\rm bulge}}$ is chosen to
be the total mass of the second spatial region and not the total mass
of the spheroid surrounding the SMBH. The binary fraction is unity,
i.e. all the bulge mass is in binaries. The total mass of each binary
is $2{\rm M_{\odot}}$ with equal mass components in a circular orbit.
The distribution of sma of the binaries is log uniform with $a\in\left\{ a_{{\rm min}},a_{{\rm max}}\right\} $
with $a_{{\rm min}}=0.01{\rm AU}$ and $a_{{\rm max}}=100{\rm AU}$.
The initial velocity dispersion is $\sigma_{0}=67.5{\rm kms^{-1}}$
and the fraction of TDE as a function of sma is taken from (\ref{eq:f_TDE}).
The initial value of the velocity dispersion is smaller than the predicted
value from the $M_{\bullet}-\sigma$ relation, almost by a factor
of two. Figure (\ref{fig:Example_time_evolution}) present the time
evolution of the velocity dispersion and the mass of the SMBH. The
SMBH accretes small amount of mass, roughly $1\%$ of its initial
mass. However the velocity dispersion is changing significantly by
almost a factor of two. The final value of $\sigma$ agrees well with
the observed $M_{\bullet}-\sigma$ relation. This simulation shows
promise and so we undertake a thorough exploration of parameter space. 

\subsection{Main results}
\label{subsec:Main-results}
We consider two binary sma distributions: log-uniform and log-normal
\citep{Duchene2013}. For the log-uniform case we assume the binary
sma are distributed from $a\in\left[a_{{\rm min}},a_{{\rm max}}\right]$,
and $a_{{\rm min}}=0.01$ and $a_{{\rm max}}=100{\rm AU}$ where the
lower bound corresponds to contact binaries for a binary with two
$1{\rm M_{\odot}}$ components. Furthermore, we assume that the mass
of the surrounding spheroid is $M_{{\rm bulge}}=\alpha\times M_{\bullet,0}$
where $M_{\bullet,0}$ is the initial mass of the SMBH, and $\alpha=50.$
The 10 initial values of the SMBH masses and 10 values of the velocity
dispersion are chosen to be evenly distributed in $\log M_{\bullet}$
and $\log\sigma$. The mass boundaries are $M_{{\rm \bullet min}}=10^{5}{\rm M_{\odot}}$
and $M_{{\rm \bullet max}}=10^{8}{\rm M_{\odot}}$ while the velocity
dispersion boundaries are $\sigma_{{\rm 0min}}=30{\rm kms^{-1}}$
and $\sigma_{{\rm 0max}}=120{\rm kms^{-1}}$.

\begin{figure}
\includegraphics[width=0.9\columnwidth]{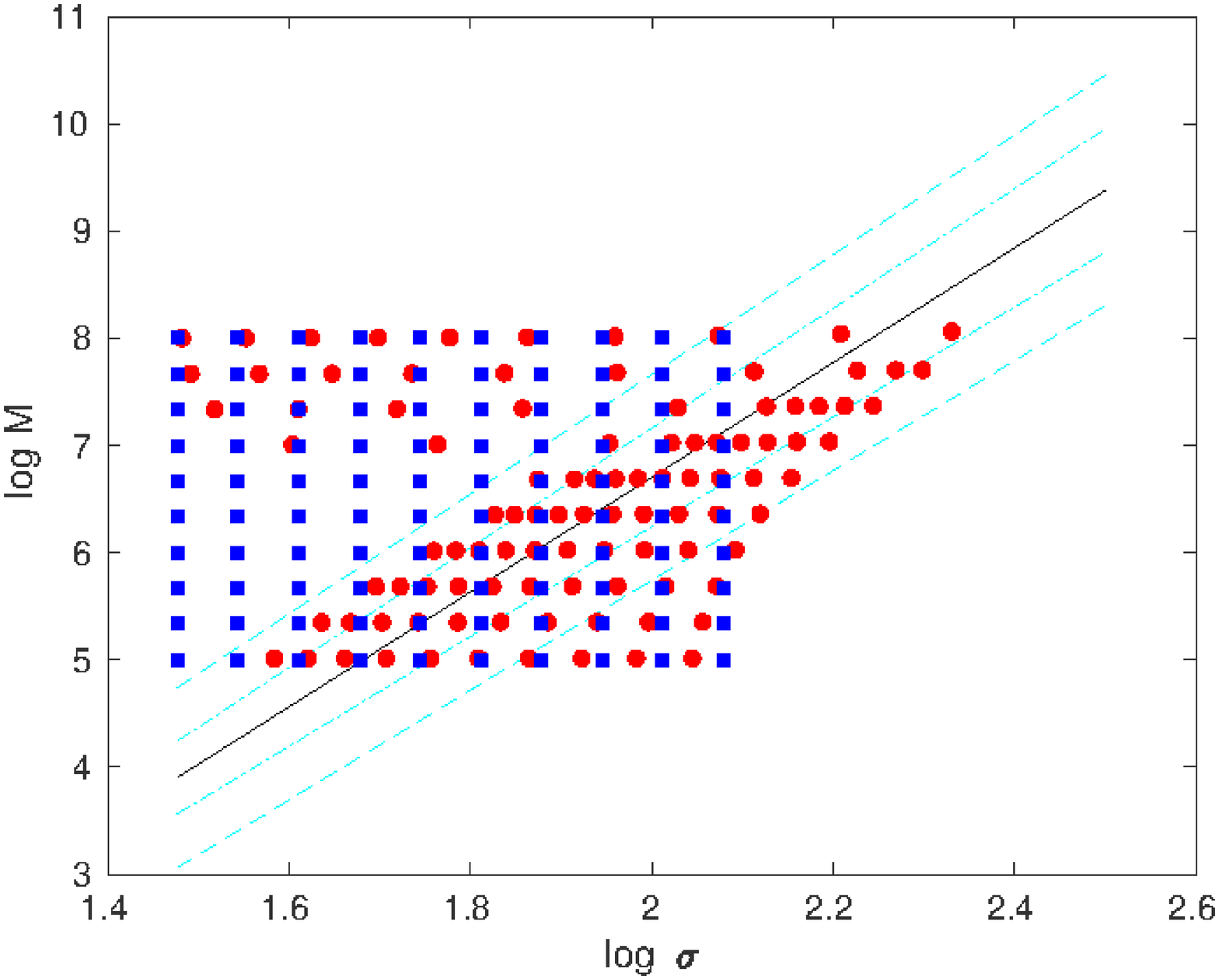}

\includegraphics[width=0.9\columnwidth]{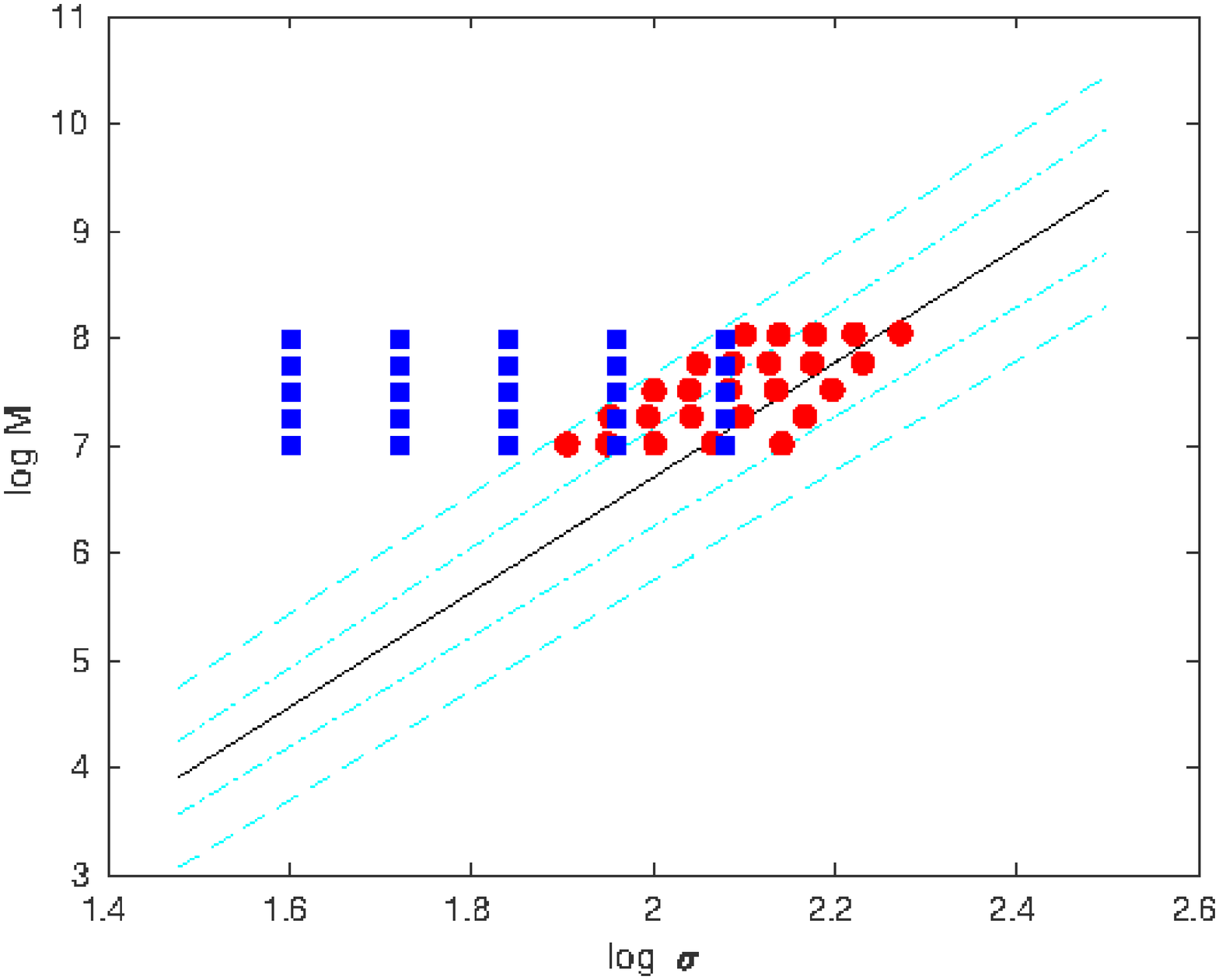}\caption{\label{fig:Log-uniform-case}Upper plot: Log-uniform distribution
of binary separations for 100 different initial conditions. Blue squares
are the initial conditions. Red circles are the values after integration
time of $t_{{\rm final}}=1.2\times10^{10}{\rm yr}$. The binary fraction
we use is $f_{{\rm binary}}=1/2$, $a_{{\rm min}}=0.01{\rm AU},$
$a_{{\rm max}}=100{\rm AU}$ and $\alpha=50$. The black solid line
is the observed $M_{\bullet}-\sigma$ relation taken from \citet{vandenBosch2016}
and the 4 cyan lines indicates sigma uncertainty in slope and intercept.
Lower plot: same as left plot with $t_{{\rm final}}=10^{12}{\rm yr}.$}

\end{figure}

Figure \ref{fig:Log-uniform-case} presents the results of 100 integrations
with $t_{{\rm final}}=1.2\times10^{10}{\rm yr}$ that corresponds
to $z=0.1$. It is clear that almost all initial condition within
the mass range of $5<\log M_{\bullet}<7$ evolve into the region were
the $M_{\bullet}-\sigma$ is observed. However, the upper range of
masses, $\log M_{\bullet}>7$ evolve into the observed region only
for sufficiently high initial velocity dispersion, i.e. $\sigma_{0}\gtrsim80{\rm kms^{-1}}$.
In order to demonstrate that the convergence does occur for the larger
masses but on only on unphysical timescale we present in the right
plot of figure \ref{fig:Log-uniform-case} the results of 25 integration
with $t_{{\rm final}}=10^{12}{\rm yr}$.

\begin{figure}
\includegraphics[width=0.9\columnwidth]{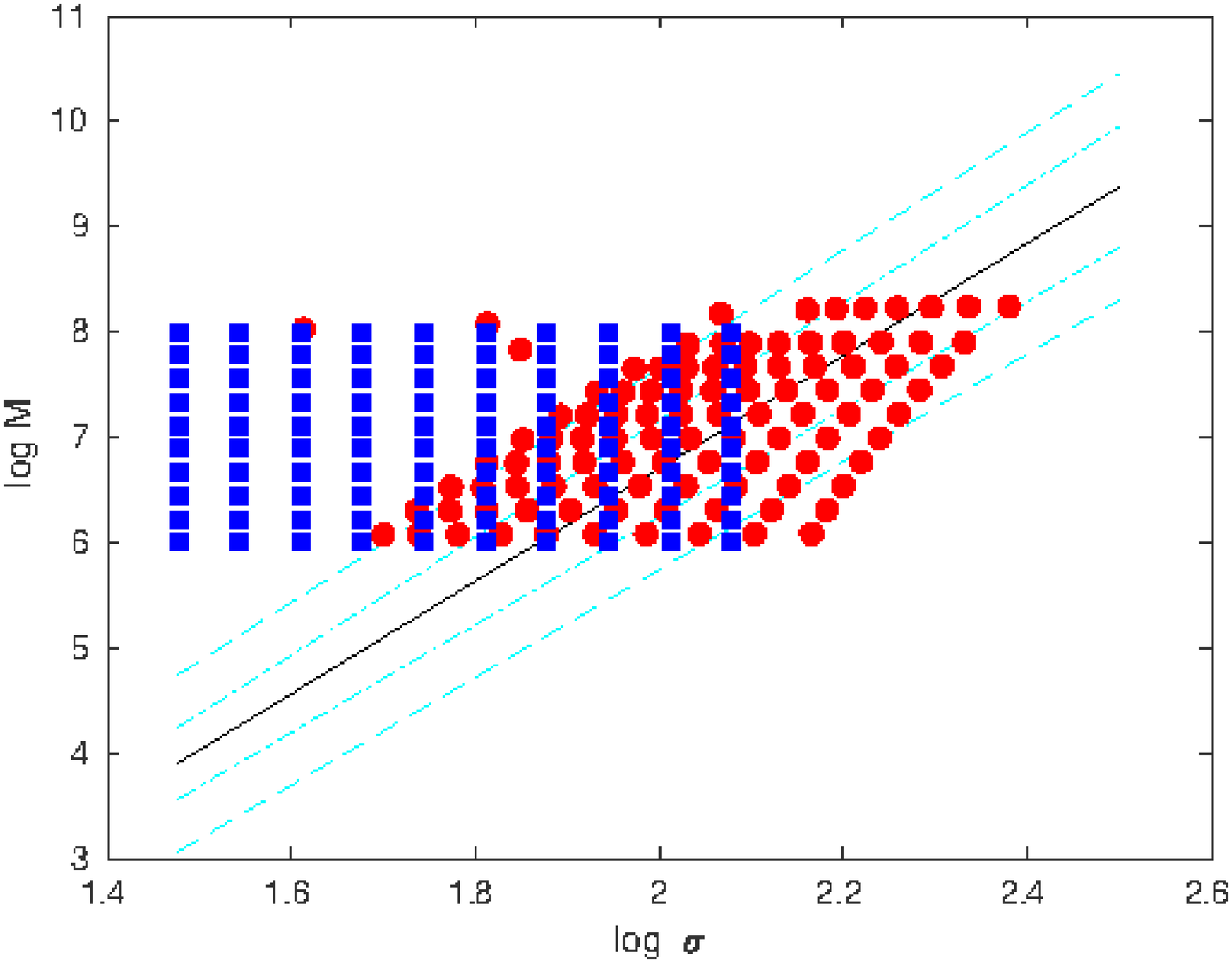}\caption{\label{fig:Log-normal}Log-normal distribution of binary separations.
100 different initial conditions. Here $a_{{\rm mean}}=60{\rm AU}$.
Blue squares are the initial conditions. Red circles are the values
after integration time of $t_{{\rm final}}=1.2\times10^{10}{\rm yr}$.
The binary fraction we use is $f_{{\rm binary}}=1/2$, $a_{{\rm min}}=0.01{\rm AU},$
$a_{{\rm max}}=100{\rm AU}$ and $\alpha=50$. The black solid line
is the observed $M_{\bullet}-\sigma$ relation taken from \citet{vandenBosch2016}
and the cyan lines are in figure \ref{fig:Log-uniform-case}.}

\end{figure}

Next we assume that the binaries sma have a log-normal distribution
with mean value of $a_{{\rm mean}}=60{\rm AU}$ \citep{Duchene2013}.
The lower bound of the sma is $a_{{\rm min}}=0.01{\rm AU}$ and the
upper bound is $a_{{\rm max}}=100{\rm AU}$. The initial conditions
are chosen to be identical as the previous case. The results are presented
in figure \ref{fig:Log-normal}. Similar to the log uniform case,
the evolution of the initial condition mimics the observed $M_{\bullet}-\sigma$
relation for the mass range $5<\log M_{\bullet}<7$ for all calculated
velocity dispersions, while for the initially more massive SMBH only
velocity dispersion greater than $\sigma_{0}\gtrsim40{\rm kms^{-1}}$.

\begin{figure*}
\includegraphics[width=0.95\columnwidth]{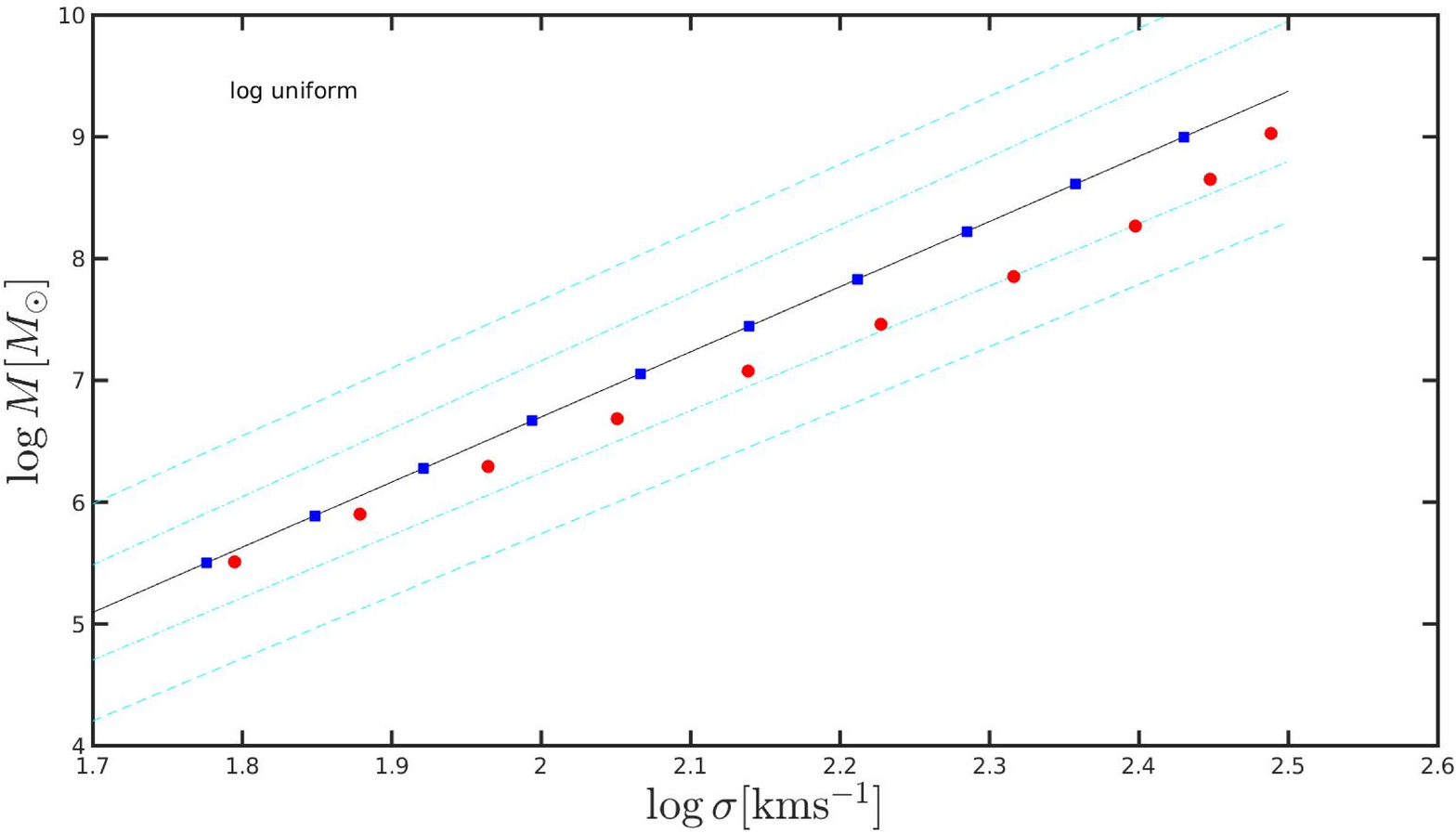}\includegraphics[width=0.95\columnwidth]{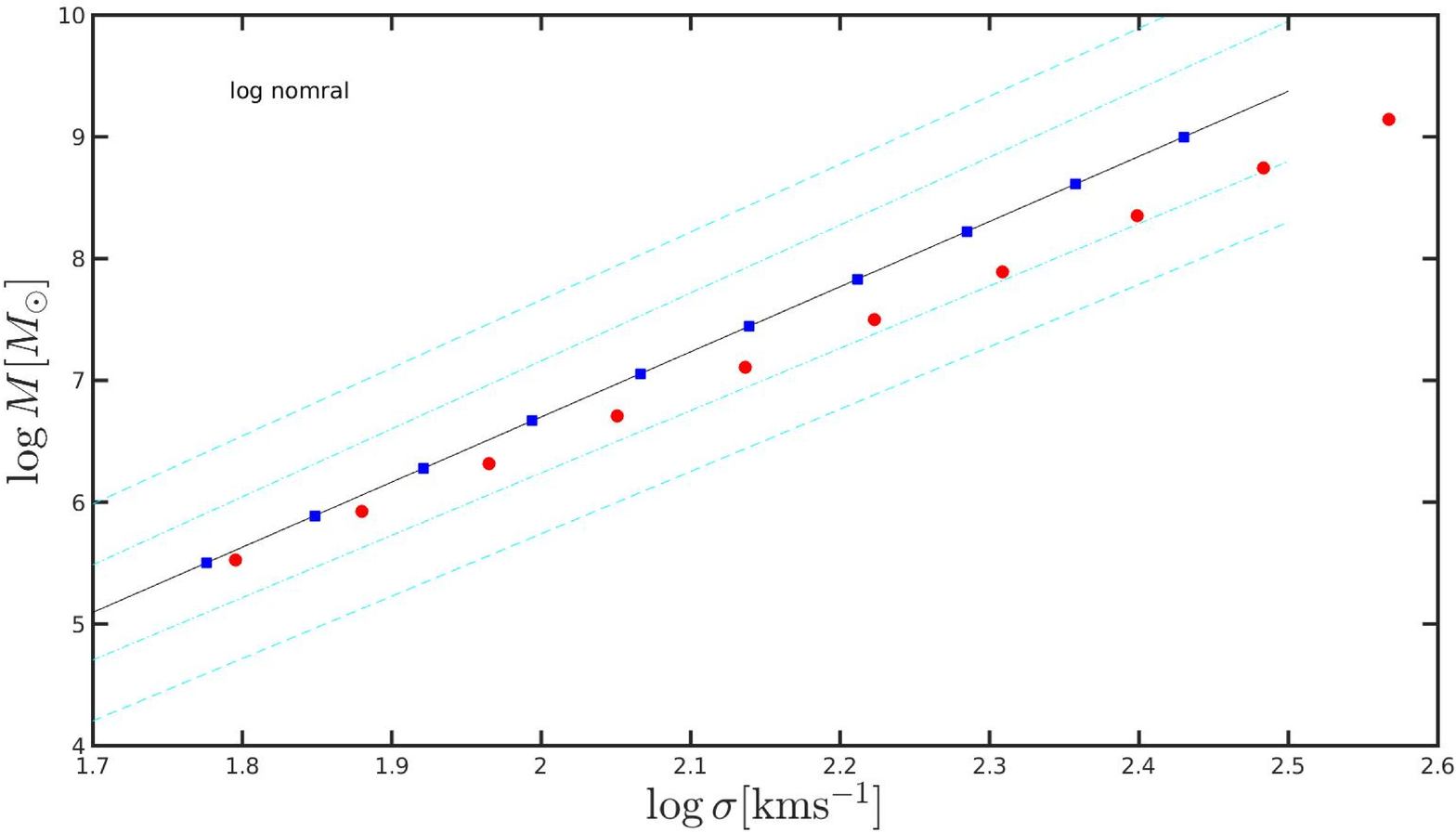}\caption{\label{fig:Stability-check}Left panel: For $\log$-uniform sma distribution.
Blue squares are the initial conditions on the best fit observed $M_{\bullet}-\sigma$
relation. The galaxy properties are the same as in figure \ref{fig:Log-uniform-case}.
Red circles are the values after $t_{{\rm final}}=1.2\times10^{10}{\rm yr}$.
Right panel: For log-normal case, galaxy properties the same as in
figure \ref{fig:Log-normal}.}

\end{figure*}

Next we checked the stability of the $M_{\bullet}-\sigma$ relation.
We set the initial conditions to be exactly on the observed relation
and checked whether the binary disruption process destroys the $M_{\bullet}-\sigma$
relation. Figure \ref{fig:Stability-check} shows the results for
both sma distributions.

\section{Discussion and Summary}
\label{sec:Discussion-and-Summary}
\subsection*{Relaxation time and radial orbits}

Our model is based on the assumption that the binary component which
returns to the bulge immediately equilibrates its excess energy with
its environment. In what follows we justify this assumption for lower
end of the SMBH masses. The star that returns to the bulge interacts
with its environment primarily via two-body interactions. Therefore
the equilibrium timescale is the two-body relaxation timescale which
is given by: 
\begin{equation}
t_{{\rm relax}}\approx\frac{\sigma^{3}}{8\pi G^{2}m^{2}n\ln\Lambda}
\end{equation}
where $m$ is the mean mass of the components, $n$ is the number
density and the Coulomb logarithm, $\ln\Lambda$, is defined as the
natural log of the ratio between the two relevant length scales of
the problem, the size of the environment, $R$, and the mean distance
between the stars, $\sim n^{-1/3}$. The spatial region we focus on,
i.e. the intermediate region (PM1), has a radius of $R\approx\alpha/2\times r_{h}$
and a mass of $\sim\alpha\times M_{\bullet}$. Hence, for a $M_{\bullet}=10^{6}{\rm M_{\odot}}$,
$\alpha=50$, and $\sigma=50{\rm kms^{-1}}$the relaxation time is
\begin{equation}
t_{{\rm relax}}\approx\frac{\alpha^{2}GM_{\bullet}^{2}}{48\sigma^{3}\bar{m}\ln\Lambda}\approx3\times10^{9}{\rm yr}\left(\frac{M_{\bullet}}{10^{5}M_{\odot}}\right)^{2}\left(\frac{\sigma}{50{\rm kms^{-1}}}\right)^{-3}
\end{equation}
which is shorter than the Hubble time. However, the relaxation time
for more massive SMBH are longer than Hubble time, therefore some
stars may not equilibrate their excess energy with the bulge. These
stars have higher speeds and move on an almost radial trajectories.

Binary disruptions that occur sufficiently late have no time to equilibrate
their excess energy with the environment. Therefore, the surviving
star will retain its velocity and have a radial trajectory within
the bulge. As a result it will have distinctively different velocity
than its neighboring stars. Can this be tested observationally?

This might be a challenge for observers when attempting to determine
the velocity dispersion from the width of some spectral lines. The
current procedure assumes a homogeneous ensemble of stars, in contrast
of what we predict here. A detail treatment of their implications
on observational signatures will be the object of future work.

\subsection*{TDE rates}

\begin{figure}
\includegraphics[width=0.9\columnwidth]{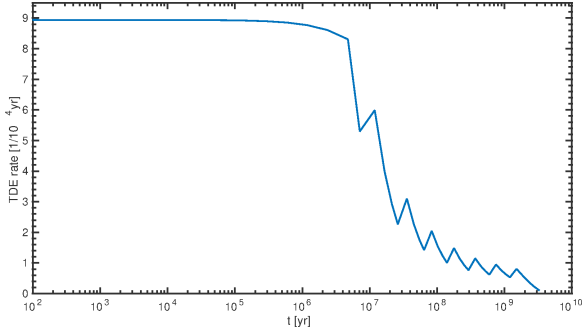}\caption{\label{fig:TDE_rate}$\alpha=50$, $M_{\bullet}=1\times10^{7}{\rm M_{\odot}}$,
$\sigma_{0}=70{\rm kms^{-1}}$, $a_{{\rm min}}=0.01{\rm AU},$ $a_{{\rm max}}=100{\rm AU},$
log-uniform distribution of the sma, $f_{{\rm TDE}}=0.1\times\left(\frac{a}{{\rm AU}}\right)^{-0.2244}$,
and binary fraction of unity, $f_{{\rm binary}}=1.$ TDEs rates are
roughly constant for the first $10{\rm Myr}$ and decrease for the
next $\sim{\rm 1Gyr}$, as the binary population dwindles.}

\end{figure}

The proposed process produces a novel channel for TDEs, namely TDEs
from binaries that originate from outside the sphere of influence.
In figure \ref{fig:TDE_rate} we present the TDE events per $10^{4}$
years for a typical galaxy parameters, see caption. However, in figure
\ref{fig:TDE_rate} we assume a single star burst of single mass components,
$m_{*}=1M_{\odot}$. We assume all stars are in their main-sequence
stage of stellar evolution, hence have the same stellar radius, $R_{*}$.
In order to study the implications of the proposed mechanism on the
TDE rates the assumptions need to be relaxed. Therefore, we reserve
a detailed analysis for the implications of TDEs rates and masses
and stellar evolution for a future work.

\subsection*{Summary}

We describe a novel explanation for the $M_{\bullet}-\sigma$ relation.
In a triaxial potential, the intermediate spatial region, from the
radius of influence, $r_{h}$, up to $\sim50\times r_{h}$, from galactic
centers, hosts chaotic trajectories of binaries. These binaries wander
sufficiently close to the center of the potential in order for the
binary to be tidally disrupted. As a result, one component loses energy
and is usually captured by the SMBH or disrupted by it, while the
other component gains energy and returns to the surrounding spheroid
with excess energy. The excess energy may be equilibrated with the
environment within a two-body relaxation time scale and hence changes
the velocity dispersion. Essentially the SMBH splits the binaries
and frees latent orbital energy and changes the kinetic temperature
of the surroinding spheroid. The change in the kinetic temperature
is equivilant to a change in the velocity dispersion that converges
to the observed $M_{\bullet}-\sigma$ relation.

Our results are robust for SMBH masses $M_{\bullet}<10^{7}M_{\odot}$,
namely galaxies with bulge masses lower than $10^{7}M_{\odot}$ converge
to the $M_{\bullet}-\sigma$ relation sufficiently fast. Accounting
for galaxy evolution theory that suggests massive galaxies are build
from the mergers of lower mass galaxies. The proposed mechanism described
here indicates that lower mass galaxies are merging already close
to their $M_{\bullet}-\sigma$ relation values and hence creates the
merged, more massive galaxy, closer to its $M_{\bullet}-\sigma$ value.

\section*{Acknowledgements}

EM thanks Richard Mushotzky and Coleman Miller for helpful and enlightening discussions.

\end{document}